\documentclass[a4paper,11pt]{article}
\usepackage{jinstpub}
\usepackage{lineno}
\usepackage{wasysym}
\usepackage{xfrac}
\usepackage{subcaption}
\usepackage{siunitx}

\title{\boldmath Characterization of an Optical Time Projection Chamber prototype with Ar:CF$_4$ at atmospheric pressure}

\author[a,1]{R. Felkai,\note{Corresponding author.}}
\author[a]{M. Borysova,}
\author[a]{L. Moleri,}
\author[a]{J. Pienaar,}
\author[a]{D. Vartsky,}
\author[a]{A. Breskin,}
\author[b]{I. Mor,}
\author[b]{L. Weissman}
\author[a]{and S. Bressler}
\affiliation[a]{Department of Particle Physics and Astrophysics, Weizmann Institute of Science,\\Herzl St 234, Rehovot, Israel}
\affiliation[b]{Soreq Nuclear Research Center,\\Yavne, Israel}

\emailAdd{ryan.felkai@weizmann.ac.il}

\abstract{We report on the characterization of an Optical Time Projection Chamber (OTPC) developed as a prototype for neutron-induced reaction studies at Soreq Applied Research Accelerator Facility (SARAF).
The prototype was operated with Ar:CF$_4$ (95:5) at 1~bar, using parallel mesh electrodes for charge multiplication and electroluminescence, and alpha particles from a $^{241}$Am source.
Both induced charge and light signals were recorded.
A systematic characterization was carried out to study charge multiplication and scintillation yield.
We present the results of these measurements in single-gap amplification configurations, along with two-gap operation and a demonstration of synchronized optical readout.}


\begin{document}
\maketitle
\flushbottom

\section{Introduction}
\label{sec:Intro}

Neutrons are at the center of many fundamental and applied research domains, including astrophysics, nuclear energy, medical isotope production and nuclear non-proliferation.
To make progress in these areas, precise experimental data are needed, in particular cross sections of neutron-induced reactions.
However, gaps still exist in nuclear data libraries and precision measurements are needed~\cite{Dupont2020}.
The Soreq Applied Research Accelerator Facility (SARAF) Phase II~\cite{Mardor2022}, presently in construction, will provide high-intensity, pulsed, fast neutron beams particularly suited for energy-dependent measurements and studies of neutron-induced reactions on various targets.
Such an experimental program would benefit from  a versatile detector providing near-4$\pi$ acceptance, event-by-event track reconstruction and particle identification - all useful for event selection and background rejection.
Optical Time Projection Chambers (OTPCs) have already been used successfully in the context of nuclear reaction studies~\cite{Miernik2007, Gai2010, Gai2020}, and recent progress in low-noise optical sensors and fast cameras makes them an attractive solution for future experiments.

The present work is part of a project aiming to build an OTPC as a multi-purpose detector for the upgraded SARAF neutron beam.
We present here the results from a small OTPC prototype, $\sim$0.4~L of active volume,  used as a test-bench for studying detector-response, gas properties, and optical-readout development.
As a first step, the prototype was operated with an Ar:CF$_4$
(95:5) mixture at 1~bar to detect $\sim$5.5~MeV alpha particle tracks from a $^{241}$Am source ($\sim$3~cm range in our gas mixture).
The results include measurements of primary charge collection, charge multiplication and secondary scintillation yield in single-gap and two-gap mesh configurations.
Synchronized optical imaging of tracks is also demonstrated. 
These results serve as a baseline for guiding the design of a future engineering prototype~\cite{Felkai2026} that will be operated at the SARAF-II neutron beam.

This paper is organized as follows.
Section~\ref{sec:Detector} describes the experimental setup: the TPC detector, gas system, high voltage and sensors.
Section~\ref{sec:ExpProc} covers the measurement configurations, signal processing, and primary charge estimation used for gain normalization.
Section~\ref{sec:Results} compiles the results on charge multiplication, light yield in the different configurations, and preliminary optical imaging.
Section~\ref{sec:Final} discusses the results and limitations, and highlights future developments.

\subsection{OTPC working principle}
\label{subsec:DetPrinciple}

The Time Projection Chamber (TPC) concept allows three-dimensional imaging of ionization tracks deposited by charged particles in a sensitive volume~\cite{Hilke2010}.
This sensitive volume, typically filled with gas, is enclosed between a cathode and an anode and surrounded by a field cage that produces a uniform electric field.
Primary ionization electrons drift toward the anode where they produce a signal either by direct collection or through a secondary process.
The latter can be obtained in a high-field region, near the anode, by charge multiplication, electroluminescence, or both.
TPCs using electroluminescence are commonly referred to as Optical TPCs (OTPCs).

In such a device, the three-dimensional reconstruction of the primary ionization track is achieved by combining the spatial distribution of the electroluminescence light with its time evolution.
While the transverse position information of the track is directly encoded in the optical image, the longitudinal position information is reconstructed with the temporal information and the electron drift velocity.
Depending on the readout technology used, the optical signal may provide time-resolved 2D imaging (e.g. Timepix-based camera~\cite{Lowe2020}) or time-integrated images (e.g. CMOS cameras~\cite{Miernik2007, Gai2010}).
In the second case, the time information can be measured independently by a photomultiplier (PMT).

Moreover, with a reference time of the initial interaction, $t_0$, one can also reconstruct its absolute position along the drift axis.
This reference time is typically provided by the primary scintillation light, denoted S1, while the electroluminescence signal, also called the delayed secondary scintillation, is denoted S2.

\subsection{Charge multiplication and scintillation}
\label{subsec:ChargeAndScint}

Different electrode geometries can be used to obtain the high electric fields required to produce charge multiplication and the secondary scintillation, such as Gas Electron Multipliers (GEMs)~\cite{Araujo2023, Amaro2023} and Thick GEMs (THGEMs)~\cite{Lowe2020}.
A traditional choice, however, consists of a stack of parallel meshes with typical gaps of the order of millimeters~\cite{Breskin1979, Breskin1990, Deisting2021}.
The main advantage of meshes in OTPCs is their transparency to the isotropically emitted secondary scintillation.
This enables more flexibility in sensor placement while maximizing the light collection efficiency.
Moreover, this geometry produces a uniform amplification field, allowing the charge transport properties in the gas to be studied directly as a function of the field.

\subsection{Ar:CF$_4$ as a detection medium}
\label{subsec:ArCF4}

OTPCs rely on the scintillation properties of the working gas and on precise knowledge of its charge transport parameters.
These parameters determine how well the spatial information of the primary energy deposition is preserved during electron drift.

Tetrafluoromethane (CF$_4$), despite being a potent greenhouse gas, is attractive for OTPCs because it combines fast electron transport~\cite{Peisert1984, Christophorou1979} with intense scintillation extending from the UV into the visible range~\cite{Pansky1995, Brunbauer2018, Brunbauer2025}.
This spectral range is well matched to fused silica windows and to the sensitivity of many camera sensors.
Recent studies~\cite{Brunbauer2025, Amedo2023} suggest that both dissociative electronic excitation and dissociative ionization of CF$_4$ result in scintillation light: the former is commonly associated with the visible radiative de-excitation of CF$_3^*$ fragments, while the latter is associated mainly with the UV emission resulting from the de-excitation of dissociative ionization fragments.

A distinctive feature of CF$_4$ is that its secondary scintillation is closely coupled to avalanche multiplication.
In noble gases, proportional electroluminescence can occur for reduced fields, $E/N$, in a range where electronic excitations happen while ionization remains negligible or absent~\cite{Monteiro2007, Monteiro2008, Mano2022}.
Such a regime has not been observed in CF$_4$~\cite{Pansky1995}.

\begin{figure}[htbp]
    \centering
    \includegraphics[width=0.8\linewidth]{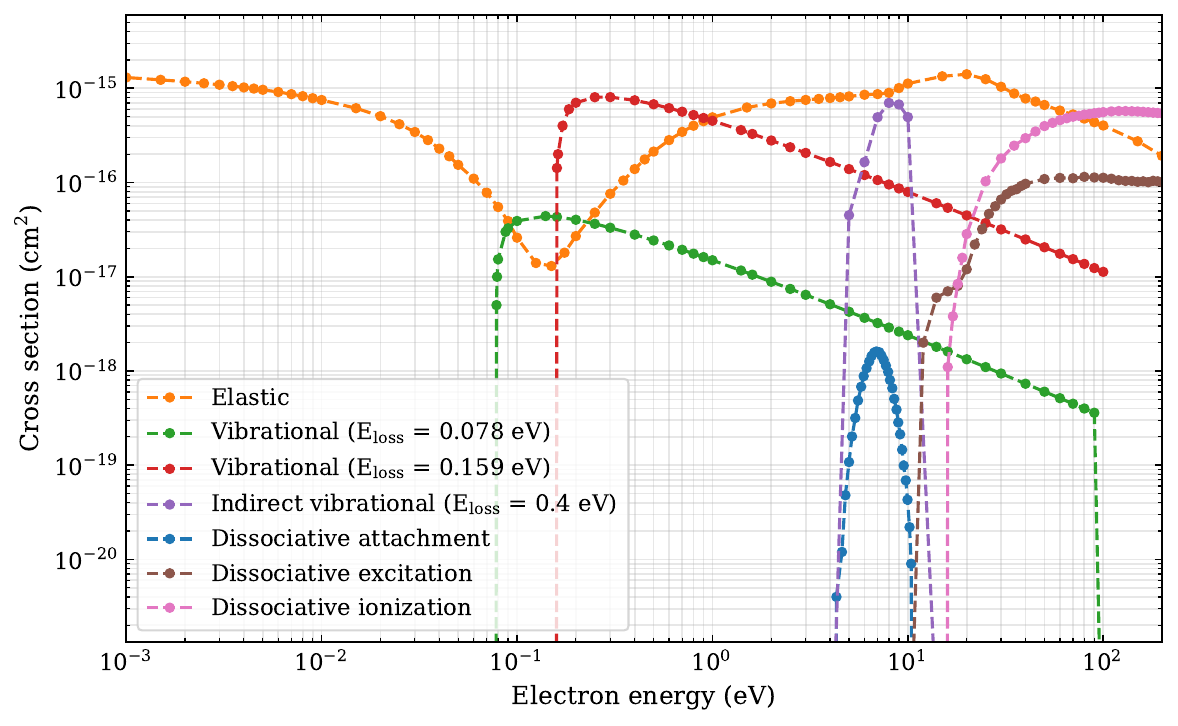}
    \caption{Electron collision cross sections in CF$_4$ from the Bordage database~\cite{Bordage1996,Bordage1999}, www.lxcat.net~\cite{LXCat2017}, retrieved on June 18, 2026.}
    \label{fig:x-sec}
\end{figure}

This behavior can be understood qualitatively from the competition between the different electron collision processes in CF$_4$, which is illustrated by the corresponding energy-dependent cross sections shown in figure~\ref{fig:x-sec}.
At low electron energies, vibrational excitation efficiently cools the electron swarm.
At intermediate energies, below the electronic excitation and ionization thresholds and above 4~eV, dissociative attachment removes electrons from the drifting swarm.
At high enough reduced fields, the high-energy tail of the electron energy distribution reaches the electronic excitation and ionization thresholds.
The competition between electron multiplication and electron loss is visible at the swarm level~\cite{Datskos1992, Anderson1992}: measurements of the effective ionization coefficient, $\bar{\alpha} = \alpha - \eta$, where $\alpha$ and $\eta$ are the ionization and attachment coefficients respectively, show negative values in CF$_4$ and CF$_4$-based mixtures over a specific range of reduced fields.
This indicates that attachment can dominate over ionization before net avalanche multiplication develops, while the proximity of the excitation and ionization thresholds means that a measurable secondary scintillation signal tends to occur under conditions where charge multiplication is also significant.

The addition of argon modifies the scintillation mechanism of the gas.
Excited argon atoms can transfer their energy to CF$_4$ molecules, effectively shifting the emitted light from the argon VUV scintillation (peaked at 128~nm) toward the CF$_4$ characteristic emission bands.
This transfer is highly efficient, even at percent-level CF$_4$ concentrations, and enhances the visible band tied to the CF$_3^*$ fragment scintillation~\cite{Amedo2023}.
Measurements of secondary scintillation spectra show that, unlike pure CF$_4$ and He:CF$_4$ mixtures, Ar:CF$_4$ mixtures maintain a dominant visible emission component~\cite{Fraga2003, Brunbauer2025}.
Moreover, an argon-dominated gas relaxes the high voltage constraints of pure CF$_4$, enabling charge multiplication at lower fields.
Ar:CF$_4$ mixtures with low CF$_4$ concentration are particularly attractive detection media for OTPCs.

\section{Detector and instrumentation}
\label{sec:Detector}

The experimental setup comprises a TPC detector with light and charge sensors, a gas system, a high voltage power supply system and readout electronics.

\subsection{TPC detector}
\label{subsec:Det}

\begin{figure}[htbp]
    \centering
    \includegraphics[width=0.7\linewidth]{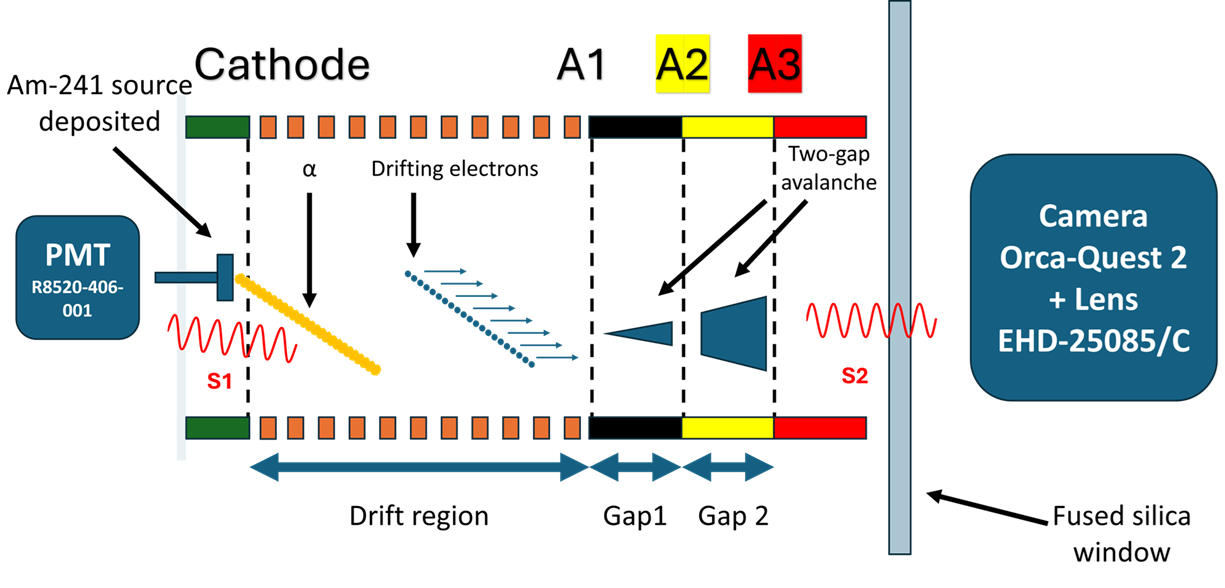}
    \caption{Schematic of the prototype showing the main components of the TPC and the photosensors layout.}
    \label{fig:tpc_schematic}
\end{figure}

The full scheme is shown in figure~\ref{fig:tpc_schematic}.
Radiation-induced ionization electrons, produced in the drift volume, drift to a first multiplication gap.
They undergo avalanche multiplication and electroluminescence photon emission.
The avalanche electrons can be transferred to a second multiplication gap for additional charge and light amplification.
Final-avalanche electrons are collected on the last electrode.
The primary scintillation photons are detected by a PMT, providing the event's time stamp. 
The electroluminescence photons are imaged by a digital camera, yielding the event's transverse coordinates (x,y) while the PMT records the time profile of the signal.

The detector consists of a small cylindrical, gas-tight stainless steel chamber (160~mm diameter and 160~mm length) housing the detector's drift volume followed by an amplification stage.
The drift region is defined by two mesh electrodes, the cathode and the first anode (A1), interconnected by a field cage.
The field cage is a stack of 21 FR4 rings with 90~mm inner diameter clad on one side by a thin copper layer.
Consecutive copper layers are separated by 2.2~mm forming a maximum drift length of 48.7~mm, between the two meshes.
To ensure the uniformity of the drift field, the cathode, copper-clad rings and A1 electrode are interconnected through a voltage divider chain of resistors, each with a resistance of 470~M$\Omega$.

The amplification stage is an implementation of the multi-step concept~\cite{Breskin1979}.
It consists of three parallel woven meshes, denoted A1, A2 and A3, which define two consecutive amplification gaps, similarly to the detector design detailed in ref.~\cite{Deisting2021}.
Each mesh, including the cathode, is made of 50~$\upmu$m diameter stainless steel wires with a 500~$\upmu$m pitch (81\% optical transparency), stretched on a circular FR4 frame with an internal diameter of 100~mm.
The gaps between meshes were defined by the mesh frames that acted as spacers: the gap spacings tested in this study were 1.6~mm, 3.0~mm and 4.6~mm.
The uniform electric fields in the two gaps (between A1 and A2 and between A2 and A3) can be adjusted independently.

In a traditional multi-step operation, e.g. with a detector using a gas mixture doped with tetrakis(dimethylamino)ethylene (TMAE)~\cite{Breskin1990, Sauvage1989}, such a configuration allows for a decoupled optimization of charge and light.
Typically, the first stages amplify the primary electrons, while the final stage operates near the ionization threshold electric field while maximizing the scintillation signal through direct excitation of TMAE by the transferred avalanche electrons.
However, as anticipated by the microphysics considerations detailed in section~\ref{subsec:ArCF4}, the scintillation mechanism in CF$_4$ cannot be untangled from ionization.
As such, light production does not occur without concurrent charge multiplication and the maximum photon yield of the detector is intrinsically limited by the maximum achievable gain.

\begin{figure}[htbp]
    \centering
    \includegraphics[width=0.7\linewidth]{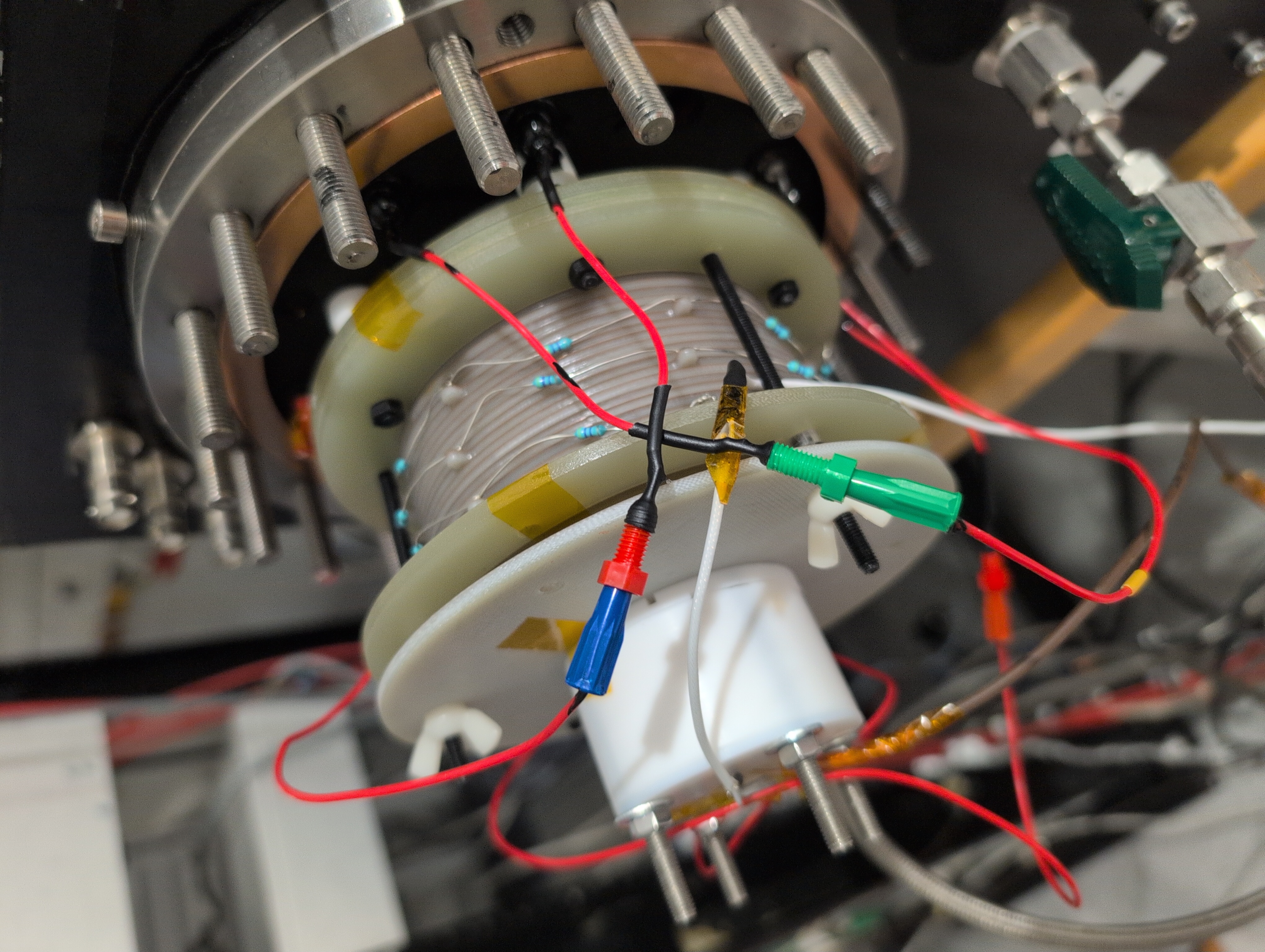}
    \caption{Picture of the fully assembled TPC on the flange. The white part of the assembly is the holder housing the PMT.}
    \label{fig:tpc_picture}
\end{figure}

A low activity ($\sim$30~Bq) Americium-241 alpha source was mounted on a transparent PMMA holder 3.0~mm behind the cathode, outside of the drift region.
Since the cathode electrode is a mesh similar to the anode ones, a large fraction of alphas enter the drift region with minimal energy loss.
The detection volume is sufficient to stop 5.5~MeV alphas emitted by the source and entering the volume.
A one-inch (25.4~mm) square UV-sensitive PMT Hamamatsu R8520-406-001 was installed within the vessel 1~cm behind the cathode, right behind the source holder.
The TPC assembly is mounted on a flange with an 80~mm diameter viewport made of Suprasil fused silica, which provides an optical interface for an external camera system.
A picture of the TPC is depicted in figure~\ref{fig:tpc_picture}.

\subsection{Gas system}
\label{subsec:Gas}

The gas system was designed to mix two gases.
The inlets were connected to high-purity (99.9999\%) Ar and (99.999\%) CF$_4$ gas cylinders.
Each line's flow was regulated by an MKS 1259-series mass flow controller (MFC) with a full-scale range of 5000~sccm and 100~sccm for Ar and CF$_4$, respectively.
The detector was operated with a relatively low total flow of 200~sccm, at a nominal Ar:CF$_4$ mixing ratio of 95:5.
The MFCs were further calibrated by a water-displacement flow measurement.
At 200~sccm, the entire gas volume is replaced in about 15~min.

The gas system and the chamber were evacuated by a turbomolecular pump.
A vacuum level of \num{1e-5}~mbar was measured directly on the vessel with an MKS 972B cold cathode gauge.
We note that this also included the outlet tube of the vessel.
The detector was then filled to reach 1.1~bar absolute pressure while keeping its outlet closed.
The detector's gas outlet was then opened to the atmosphere through a 3~m long and 3~mm inner diameter metal tube to maintain nearly atmospheric pressure within the chamber when flushing it, while suppressing air backflow.
During the filling and flushing of the chamber, the pressure was monitored with an MKS Baratron 627H capacitance manometer (1000~Torr full scale).
To ensure stable data taking conditions the detector was flushed continuously for at least 12~hours and kept under continuous flow during operation.

The gas purity was also investigated using a quantitative gas analyzer (QGA, model HAS-301-1714A from Hiden Analytical) that was installed at the outlet of the gas system.
The QGA response to the Ar and CF$_4$ components was calibrated by sampling directly a premixed Ar:CF$_4$:CO$_2$ gas cylinder (45:40:15).
We estimated from the QGA that the CF$_4$ fraction in this study was slightly above its nominal value, closer to 6\%.
Given the limited accuracy of this calibration we quote here the nominal mixing of 95:5.
The QGA also detected signals that are compatible with N$_2$, O$_2$, H$_2$, CO$_2$ and water impurities whose concentration could only be roughly estimated due to the lack of suitable calibrations.
We obtained the concentration values from comparing the strength of the impurity signals to the strength of the Ar signal and corrected them for the instrument relative sensitivity, fragmentation patterns, both provided by the manufacturer, and overlapping mass peaks.
We obtained approximately 100~ppm for N$_2$, O$_2$ and H$_2$, 10~ppm for CO$_2$ and an estimated upper limit of 300~ppm for H$_2$O.
The latter also accounts for the contribution of the doubly ionized $^{36}$Ar naturally present in $^{\mathrm{nat}}$Ar.
As the QGA samples the gas near the exhaust of the gas system, downstream of the chamber, we can qualify the quoted concentrations as upper bounds.

\subsection{High voltage, charge and light readout}
\label{subsec:HVReadout}

\begin{figure}[htbp]
    \centering
    \includegraphics[width=0.7\linewidth]{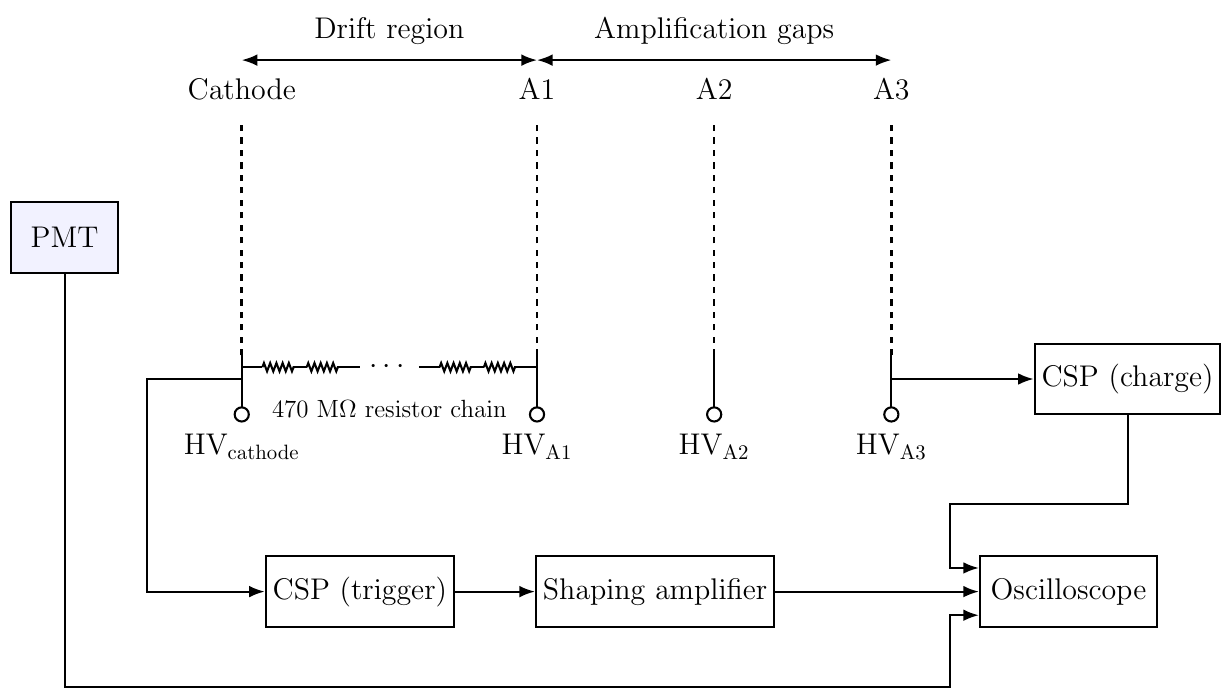}
    \caption{Schematic of the electronic readout structure and high voltage connections in the two-gap configuration.}
    \label{fig:tpc_electro_scheme}
\end{figure}

The high voltage connections and electronic readout scheme are shown in figure~\ref{fig:tpc_electro_scheme}.
Each mesh electrode was independently biased by a CAEN N1471H high voltage power supply capable of delivering up to $\pm$5~kV and 20~$\upmu$A per channel.
The bias was provided through a low-pass filter, and, for electrodes from which the signal was read out, through a charge-sensitive preamplifier (CSP) circuit.

To amplify induced charge signals we employed different CSP models, optimizing the signal-to-noise ratio while avoiding saturation.
Naicam CB200PC, featuring the highest gain of 6.2~V/pC, was used for detecting the drift of primary charges in the drift region.
Cremat CR-110, with a gain of 1.4~V/pC, was used for measurements with small or no charge amplification.
Cremat CR-112, with a lower gain of 13.5~mV/pC, was used for conditions involving substantial charge amplification.
Each Cremat CSP was mounted on a CR-150-R6 evaluation board.
The different CSP gains were determined with an independent calibration by injecting charge through the test input capacitor of each CSP (1~pF for Cremat and 3~pF for Naicam) and measuring the resulting signal amplitude.
During detector operation, a first CSP was connected to the anode at the highest voltage, where electrons were collected, and was used for signal acquisition.
A second CSP was connected to another electrode (usually the cathode) and its output was fed into a linear shaping amplifier to trigger the oscilloscope.
In the following, the term CSP refers exclusively to the signal-acquiring one.

As mentioned in section~\ref{subsec:Det}, a PMT was installed behind the cathode and the source holder.
This configuration provides high sensitivity to S1 signals, which are faint and predominantly emitted in the UV range.
A PMT calibration was performed to estimate the single-photoelectron response using the charge distribution of the PMT dark counts.
To ensure that electrons produced between the source and the detector cathode do not drift into the active volume, the detector cathode was kept at a lower voltage than the PMT photocathode, which was biased at -750~V.

All CSP and PMT anode signals were digitized and saved in binary files using the fast frame mode of a Tektronix MSO54B oscilloscope.
Waveforms from the CSP channels were acquired on a 1~M$\Omega$ input with AC coupling and a 20~MHz bandwidth limit to reduce high frequency fluctuations.
The PMT waveforms were acquired with DC coupling on a 50~$\Omega$ input and a 500~MHz bandwidth to achieve optimal timing performance.
The sampling interval was fixed at 8~ns/sample to maximize the number of waveforms recorded while retaining sufficient timing resolution.
This configuration allowed recording of 50k waveforms per run with a time window of 10~$\upmu$s or 25k waveforms per run with a time window of 20~$\upmu$s.

\section{Experimental procedure}
\label{sec:ExpProc}

\subsection{Measurement configurations}
\label{subsec:Meas}

Two different modes of operation were used: the first one involved only drifting and collection of primary electrons while the second one also involved charge multiplication and electroluminescence.

\begin{figure}[htbp]
    \centering
    \begin{subfigure}{0.48\linewidth}
        \includegraphics[width=\linewidth]{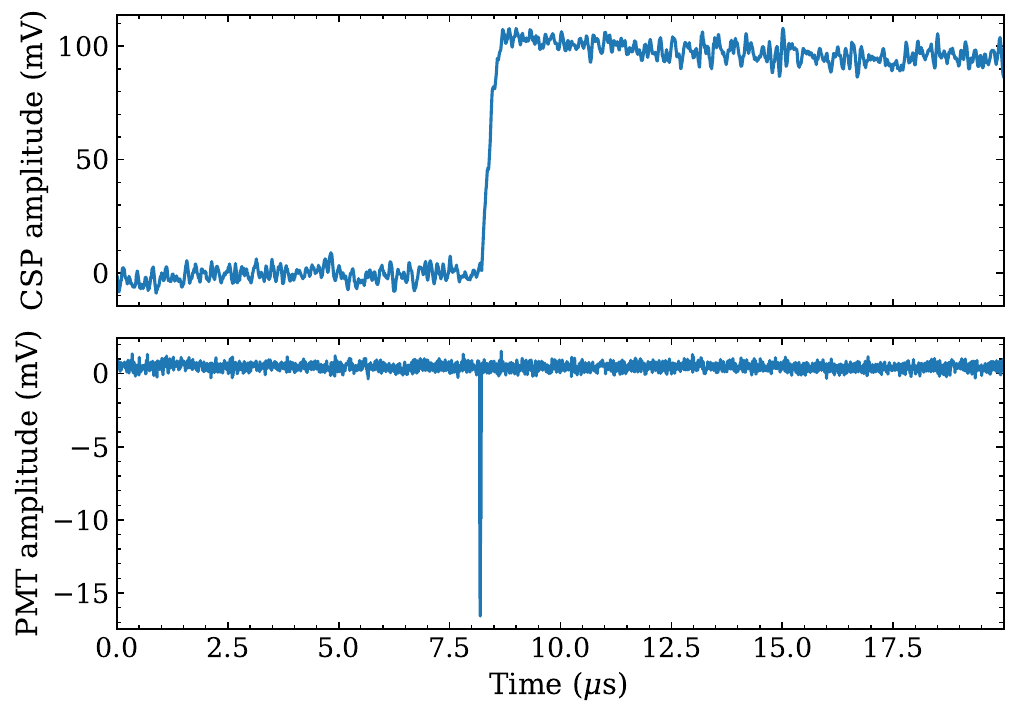}
        \caption{}
    \end{subfigure}
    \hfill
    \begin{subfigure}{0.48\linewidth}
        \includegraphics[width=\linewidth]{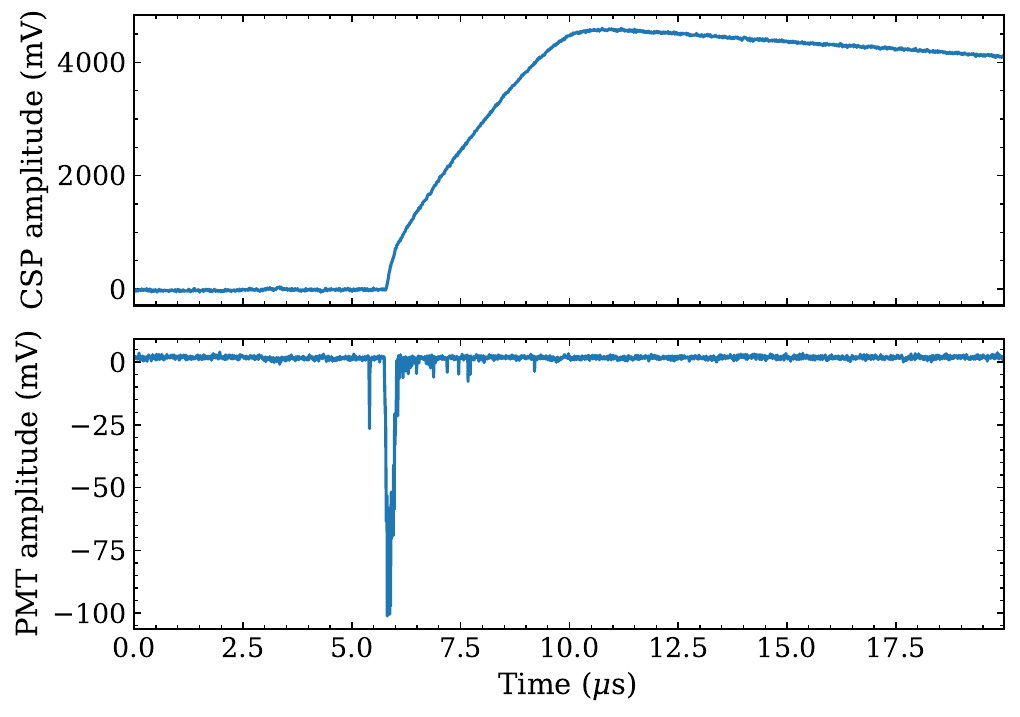}
        \caption{}
    \end{subfigure}
    \caption{Example of raw waveforms recorded. (a) In drift-only: A1 CSP signal measured with a Naicam CB200PC on the top panel and the corresponding PMT signal, including S1, on the bottom panel. (b)~With charge amplification: A2 CSP signal measured with a Naicam CB200PC on the top panel and the corresponding PMT signal, including both S1 and S2 on the bottom panel.}
    \label{fig:raw-signals}
\end{figure}

\paragraph{Primary charge drift-only}
In this configuration the primary charges released by the alpha particle interaction are drifted and collected at the first anode (A1), without charge multiplication or secondary light production.
The voltages applied to A2 and A3 were kept 50~V below that of A1, to ensure full electron collection on A1.
The voltage difference between the cathode and A1 was scanned to probe drift field values ranging from 10 to 1000~V/cm.
The resulting induced current signal was integrated by a CSP connected to A1.
An example of typical waveforms recorded in this configuration is shown on the left panel of figure~\ref{fig:raw-signals}.

In principle, the total signal amplitude provides a measurement of the primary charge, $Q_{\mathrm{primary}} = N_0\,e$ where $N_0$ is the number of ion-electron pairs produced alongside the alpha interaction.
In practice, the detector is operated without a Frisch grid and the ion component is too slow to be integrated by the CSP over the relevant time scale.
As such, and according to the Shockley-Ramo theorem~\cite{He2001}, the amplitude measured does not correspond to $Q_{\mathrm{primary}}$.
Instead, it corresponds to an electron-component effective charge, $Q_{\mathrm{primary}}^{\mathrm{eff}}$, which is dictated by the ionization profile of the alpha track and the corresponding distribution of electron drift distances.
In the following we consider this measured charge as an effective primary charge.
An estimation of the actual primary charge is done in section~\ref{subsec:SR-theorem}.

\paragraph{Charge multiplication and electroluminescence}
In this configuration the primary electrons are transferred to the amplification stage for charge multiplication and electroluminescence, either in a single-gap or two-gap configuration.
The CSP is connected to the anode on which the electrons are collected: A2 in a single-gap mode and A3 in a two-gap mode.
When electrons were collected at A2, the voltage applied to A3 was kept 50~V below that applied to A2, to ensure full electron collection.
Unlike in the previous case, the CSP is noticeably sensitive to the ion component that follows the production of avalanche electron-ion pairs.
As such, the avalanche charge, $Q_{\mathrm{avalanche}}$, was measured from the amplitude of the CSP signal and the effective gain was estimated by dividing it by the primary charge measured in drift-only operation $Q_{\mathrm{primary}}^{\mathrm{eff}}$.
An example of typical waveforms recorded in this configuration is shown on the right panel of figure~\ref{fig:raw-signals}.

From the waveforms recorded by the PMT, the following observables were measured: the scintillation time of the S1, $t_{S1}$, the start time of the S2, $t_{S2}$, the end time of the S2, $t_{S2,\mathrm{end}}$, corresponding durations and the number of photoelectrons detected $N_\mathrm{PE}$.
The procedure followed to measure these observables is described in the following section.
The absolute number of photons emitted isotropically by electroluminescence $N_{\gamma}^{4\pi}$ can be estimated from $N_\mathrm{PE}$ through the spectral efficiency of the PMT $\eta_\mathrm{spec}$ and the geometric acceptance $G$ using the following expressions:
\begin{equation}
N_{\gamma}^{4\pi} = \frac{N_{\mathrm{PE}}}{G \, \eta_{\mathrm{spec}}},
\qquad
\eta_{\mathrm{spec}} = CE \times \int S_{\mathrm{norm}}(\lambda) \, QE(\lambda) \, d\lambda,
\qquad
G = \frac{\Omega}{4\pi} \times (T_{\mathrm{mesh}})^{~n_{\mathrm{meshes}}}
\end{equation}
where $\Omega$ is the solid angle subtended by the PMT photocathode from the emission point, $n_{\mathrm{meshes}}$ is the number of meshes between the PMT and the emission point (2 in a single-gap operation and 3 in two-gap operation), $T_{\mathrm{mesh}}$ is the normal-incidence optical transmission of each mesh (81\%), $CE$ is the PMT collection efficiency (90\%), $QE(\lambda)$ is the wavelength-dependent PMT quantum efficiency, and $S_{\mathrm{norm}}(\lambda)$ is the normalized secondary scintillation emission spectrum of the gas mixture.
The PMT collection and the spectral quantum efficiencies were provided by the manufacturer and the secondary scintillation emission spectrum was taken from measurements reported in~\cite{Brunbauer2018,Brunbauer2025}.
The value calculated for $\eta_\mathrm{spec}$ is about 4.1\% while the values for $G$ are 0.58\% and 0.43\% for the single-gap and two-gap configurations, respectively.

\subsection{Signal processing}
\label{subsec:SigProc}

The charge signals recorded by the anode CSP and light signals recorded by the PMT were processed on an event-by-event basis.
A custom Python-based framework was used to read the binary oscilloscope waveform files, synchronize the recorded channels, and compute the per-event observables used in the analysis.
The resulting event tables were then stored in HDF5 summary files.
For each run, the reconstructed observables at the event level were collected into distributions.
For the charge and light observables, the main peak of the distributions was fitted with a Gaussian function.
The fitted mean was taken as the representative run level value, while the fitted standard deviation was used to estimate the event-to-event spread of the value.

\begin{figure}[tbp]
    \centering
    \begin{subfigure}{0.48\linewidth}
        \includegraphics[width=\linewidth]{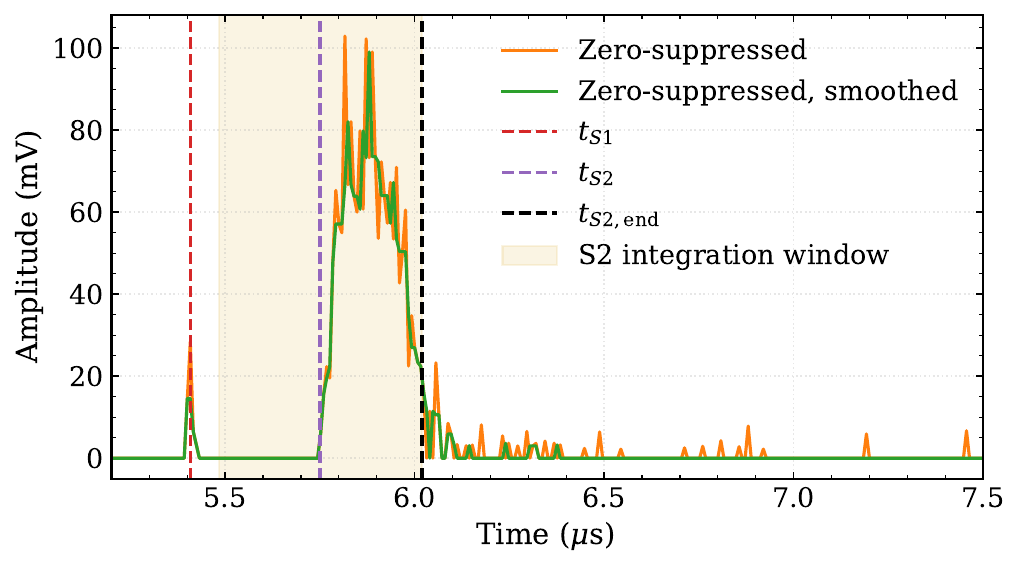}
        \caption{}
    \end{subfigure}
    \hfill
    \begin{subfigure}{0.48\linewidth}
        \includegraphics[width=\linewidth]{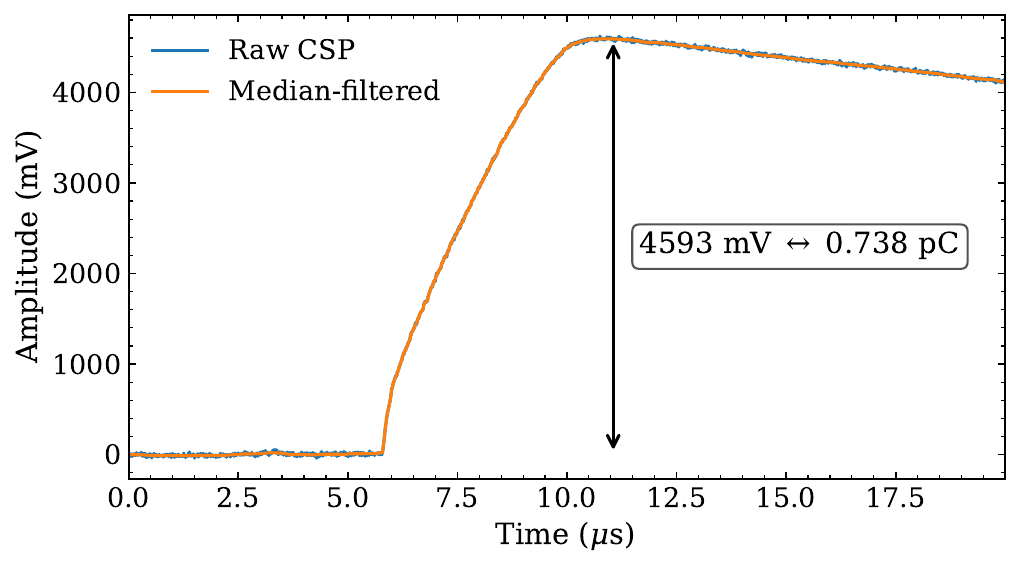} 
        \caption{}
    \end{subfigure}
    \caption{Representative waveform processing for the PMT and CSP signals. (a) PMT waveform illustrating the identification of S1, S2 and the integration window used to determine the photoelectron charge associated with S2. (b) CSP waveform from which the collected charge is obtained as the maximum signal amplitude.}
    \label{fig:waveform-processing}
\end{figure}

\paragraph{PMT signals}
An example of the PMT waveform processing is shown on the left panel of figure~\ref{fig:waveform-processing}.
The baseline offset of each PMT waveform was computed as the average of the first $1~\upmu$s of the acquisition window, corresponding to a pre-S1 region.
The baseline was then subtracted from the waveform before flipping its polarity.
The PMT baseline noise was estimated at the run level to define
a zero-suppression threshold:
\begin{equation}
    T_{\mathrm{ZS}} = \max(3.5\sigma_{\mathrm{noise\,PMT}}, 2~\mathrm{mV})
\end{equation}
where $\sigma_{\mathrm{noise\,PMT}}$ is the noise standard deviation in the pre-S1 region.
Time samples below this threshold were then set to zero. 
This noise-dependent procedure allowed the zero-suppression threshold to adapt to the increase in digitized baseline fluctuations associated with larger oscilloscope vertical scales.
The 2~mV floor threshold preserved sensitivity to single photons under low-noise conditions.

A threshold discrimination logic was applied to find S1 and S2 signals.
The S1 threshold was defined as:
\begin{equation}
    T_{S1} = \max(4~\mathrm{mV}, T_{\mathrm{ZS}})
\end{equation}
The first threshold crossing defined an S1 candidate.
The maximum amplitude within a 75~ns window starting from that crossing defined the S1 peak time, $t_{S1}$.
This S1 window duration was chosen because it is long compared with the 15~ns decay time of the visible scintillation of CF$_4$~\cite{Morozov2011} while remaining short compared to the electron drift times expected here.

The S2 start was defined as the first sample crossing a threshold $T_{S2} = 1.2 T_{S1}$ after the end of the S1 window.
This $T_{S2}$ was selected to suppress spurious single photons more strongly than for the S1 detection while remaining sensitive to the onset of the secondary scintillation.
The waveform was smoothed with a median filter using a kernel of three samples (24~ns) in order to remove the photon shot noise that sits on top of the S2 signal shape.
The reported S2 start time, $t_{S2}$, was obtained by interpolating the threshold crossing between the two adjacent time samples.
The electroluminescence delay, $\Delta t$, could be defined as $\Delta t = t_{S2} - t_{S1}$.

The S2 end time, $t_{S2,\mathrm{end}}$, was determined similarly to $t_{S2}$, from the same smoothed waveform and with the same interpolation logic.
Two different falling edge thresholds were defined: a nominal one at 20\% of the S2 maximum amplitude, and a fallback one, set to 50\% of the S2 maximum amplitude in order to mitigate the rise of a signal tail developing at the highest field configurations and attributed to feedback effects.
The condition to branch to the fallback logic was for the median PMT peak amplitude to be above 600~mV at the run level.
This was found empirically to be an appropriate value, keeping the S2 duration distribution consistent across runs taken at different amplification fields, as expected when the drift field and track geometry remain unchanged.
The S2 duration, $\Delta t_{S2}$, could then be defined as $\Delta t_{S2} = t_{S2,\mathrm{end}} - t_{S2}$.

The PMT signal photoelectron charge was computed by integrating the non-smoothed waveform from the end of the S1 window until $t_{S2,\mathrm{end}}$.
In the absence of a properly reconstructed S2, the photoelectron charge was computed until the end of the time window.
This method allows estimating the onset of light emission as well as separating the track signal from the signal tail at high fields that may have resulted from possible photon-feedback effects.
The photoelectron charge was then converted to a corresponding number of photoelectrons using the PMT calibration described in section~\ref{subsec:HVReadout}.

\paragraph{CSP signals}
CSP waveforms were used to measure the collected charge in both drift-only and amplification modes.
For the primary charge drift-only measurement described in section~\ref{subsec:Meas}, electrons were collected on A1 without charge multiplication and CSP waveforms were recorded from A1.
For measurements involving charge multiplication, the CSP waveforms were instead recorded from A2 in single-gap operation or from A3 in two-gap operation.
A representative waveform taken in charge multiplication operation is shown on the right panel of figure~\ref{fig:waveform-processing}.
The CSP baseline was computed from the waveform samples preceding $t_{S1}$, which was obtained from the PMT signal.
Similarly to the PMT case, the baseline was then subtracted from the waveform.
The CSP waveforms were smoothed with a 51-sample median filter to remove noise while preserving the monotonic rise of the signal.

The maximum amplitude of the processed waveform was converted to the collected charge using the corresponding calibration of the CSP used (see section~\ref{subsec:HVReadout} for the description of the calibration procedure).
For the reference measurement of the primary charge, this gives the effective primary charge collected on A1, $Q_{\mathrm{primary}}^{\mathrm{eff}}$, while for amplification measurements this gives the avalanche charge, $Q_{\mathrm{avalanche}}$.

\subsection{Primary charge measurement and gain normalization}
\label{subsec:SR-theorem}

\begin{figure}[htbp]
    \centering
    \begin{subfigure}{0.48\linewidth}
        \includegraphics[width=\linewidth]{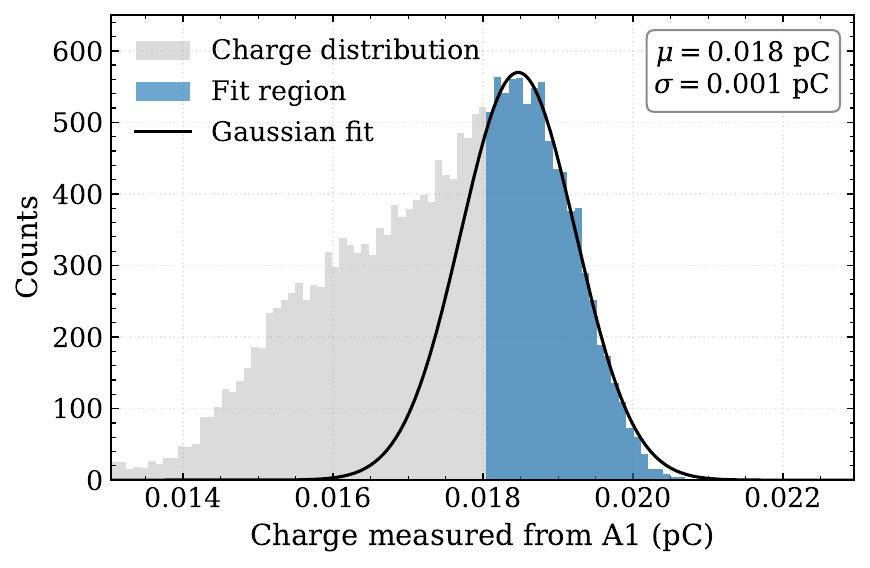}
        \caption{}
    \end{subfigure}
    \hfill
    \begin{subfigure}{0.48\linewidth}
        \includegraphics[width=\linewidth]{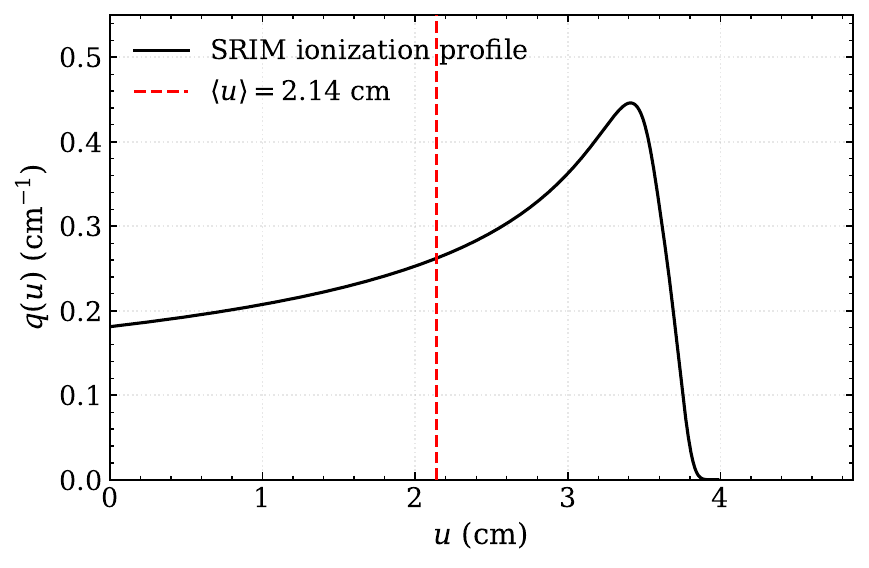} 
        \caption{}
    \end{subfigure}
    \caption{(a) Distribution of the effective primary charge measured on A1, the Gaussian fit is performed on the unattenuated peak region. The broad spectrum results from the source emission geometry. (b) Normalized ionization profile along the longitudinal dimension of a forward-emitted alpha particle obtained from a SRIM~\cite{SRIM2010} simulation. The dashed line indicates the ionization-weighted mean position, $\langle u\rangle$.}
    \label{fig:primary-normalization}
\end{figure}

The effective primary charge was measured in drift-only operation, following the procedure described in section~\ref{subsec:Meas}.
The left panel of figure~\ref{fig:primary-normalization} shows an example of the resulting charge distribution, with a Gaussian fit performed on the unattenuated peak to retrieve the representative charge.
The broad distribution results from the angle-dependent energy loss of alpha particles emitted from the source before entering the active volume.
A reference effective primary charge of 0.0184~pC was obtained from the plateau of the primary charge measurements as a function of the drift field.

As discussed in section~\ref{subsec:Meas}, the effective primary charge measured on A1 does not correspond to the total primary ionization charge.
According to the Shockley-Ramo theorem~\cite{He2001}, and assuming a uniform electric field, the charge induced by an electron produced at a distance $u$ from the cathode is proportional to its drift distance, $D-u$, where $D$ is the cathode-A1 distance.
To estimate the relation between the measured effective primary charge and the actual number of electron-ion pairs, we used the normalized longitudinal ionization profile, $q(u)$, of a forward-emitted alpha particle obtained from a SRIM~\cite{SRIM2010} simulation in Ar:CF$_4$ (95:5) at 1~bar.
The profile, which is shown on the right panel of figure~\ref{fig:primary-normalization}, was truncated by 3~mm before normalization to account for the energy lost by the alpha before entering the active volume.
The ratio between the effective and total primary charge is
\begin{equation}
    \frac{Q_{\mathrm{primary}}^{\mathrm{eff}}}{Q_{\mathrm{primary}}} = \int_0^D q(u)\frac{D-u}{D}du = \frac{\langle z\rangle}{D},
    \label{eq:shockley-ramo-correction}
\end{equation}
where $\langle z\rangle$ is the ionization-weighted mean electron drift distance.
We obtain from the ionization profile $\langle z\rangle = 2.73$~cm ($\langle u\rangle = 2.14$~cm), so the effective primary charge measured is 56\% of the total primary charge.
We can then estimate that the primary charge to be about 0.033~pC (\num{2.05e5}~primary electrons).

The effective charge gain reported in the following section is defined as $Q_{\mathrm{avalanche}}/Q_{\mathrm{primary}}^{\mathrm{eff}}$.
Under the approximation laid out above, the effective gain overestimates the gain normalized to the total primary charge by a factor of approximately 1.79.
An attempt was made to use A1 as a Frisch grid and measure the total primary charge from A2, but the balance between charge transfer efficiency through A1 and the onset of attachment made the measurement unreliable.

\section{Results}
\label{sec:Results}

\subsection{Single-gap operation}
\label{subsec:1Gap}

We have investigated three different spacings for the first gap: 1.6~mm, 3.0~mm and 4.6~mm.
For the rest of the study the drift field was fixed at 200~V/cm.

\begin{figure}[htbp]
    \centering
    \begin{subfigure}{0.48\linewidth}
        \includegraphics[width=\linewidth]{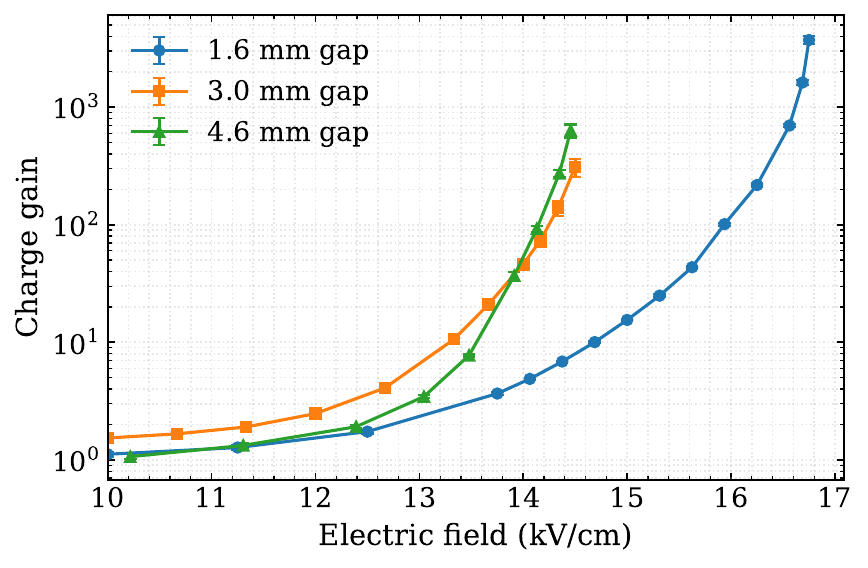}
        \caption{}
    \end{subfigure}
    \hfill
    \begin{subfigure}{0.48\linewidth}
        \includegraphics[width=\linewidth]{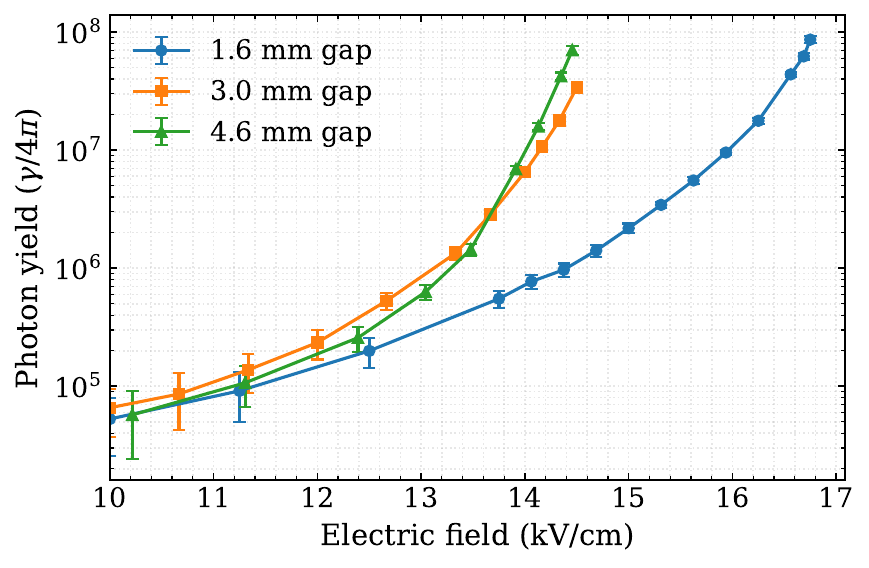}
        \caption{}
    \end{subfigure}
    \caption{(a) Effective gain curve for different gap spacings as a function of the gap electric field. (b) Inferred secondary scintillation photon yield as a function of the same electric field.}
    \label{fig:1gap-results}
\end{figure}

The primary electrons produced by the alpha particle interaction are transferred to the first gap and then drift towards the anode A2.
A2 is connected to the CSP to measure the charge following the procedure described in section~\ref{subsec:SigProc}.
The effective charge gain as a function of the field across the three different gap spacings is shown on the left panel of figure~\ref{fig:1gap-results}.
We observe a deviation from the expected exponential behavior that we attribute to feedback ionization effects.
The effective gain was calculated using the primary charge reference and normalization described in section~\ref{subsec:SR-theorem}.

\begin{figure}[htbp]
    \centering
    \begin{subfigure}{0.48\linewidth}
        \includegraphics[width=\linewidth]{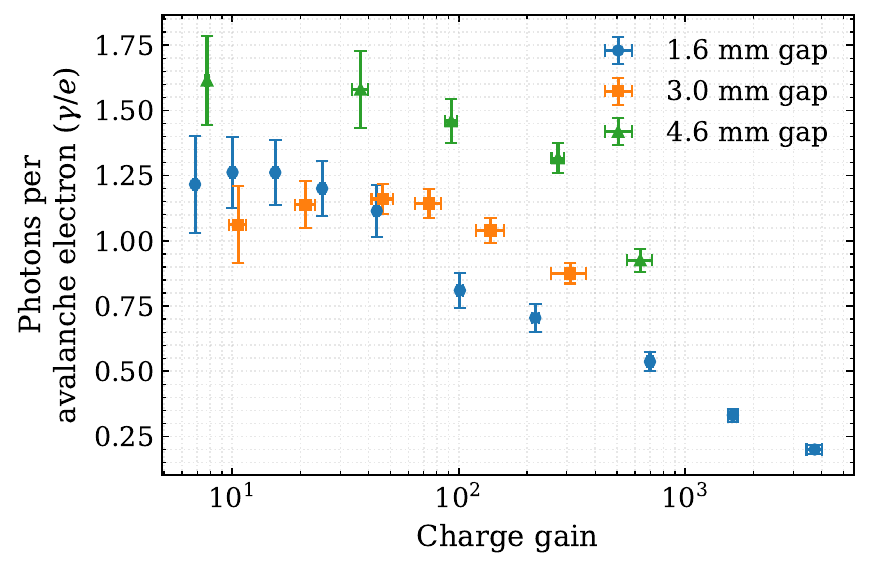}
        \caption{}
    \end{subfigure}
    \hfill
    \begin{subfigure}{0.48\linewidth}
        \includegraphics[width=\linewidth]{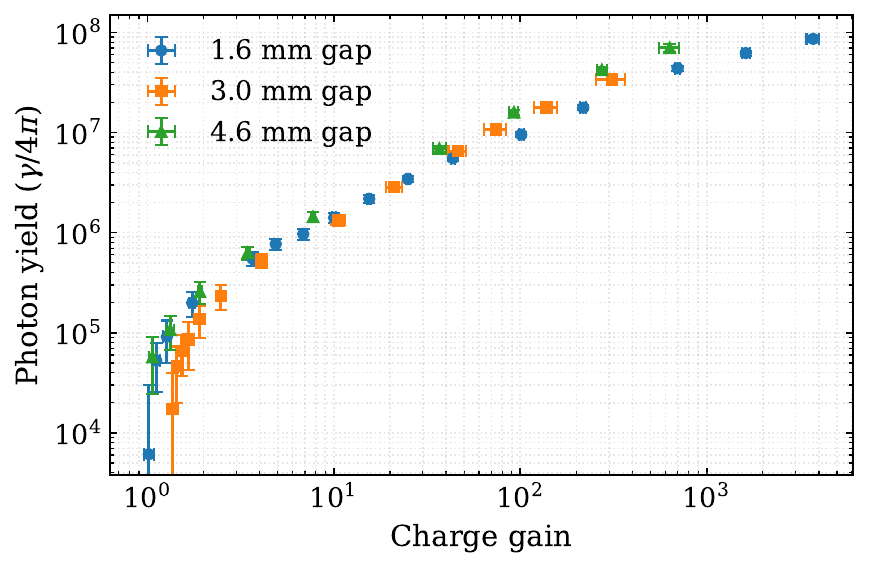}
        \caption{}
    \end{subfigure}
    \caption{(a) Relative photon yield ($\gamma$/e) as a function of the effective charge gain. The values shown only include conditions where a substantial gain is achieved. (b) Inferred photon yield ($\gamma/4\pi$) as a function of the effective charge gain.}
    \label{fig:1gap-combined-results}
\end{figure}

The absolute isotropic photon yield is inferred at the run level from the PMT waveforms recorded, following the signal-processing procedure described in section~\ref{subsec:SigProc}.
The relative photon yield (per secondary electron) is also determined at the run level from the ratio of the absolute yield to the charge obtained directly from the corresponding PMT and CSP waveforms.
The isotropic photon yield as a function of the electric field is shown on the right panel of figure~\ref{fig:1gap-results}.
Here, too, we observe a deviation from the expected exponential behavior. 
The relative photon yield as a function of the charge gain is shown on the left panel of figure~\ref{fig:1gap-combined-results}.
Because the photon yield was obtained by integrating the PMT waveforms until the reported falling edge threshold, the values reported here exclude the tail of the signal.
This tail of the signal is not usable for time-resolved imaging of the tracks.

A complementary view can be obtained by plotting the absolute photon yield as a function of the charge gain, as shown in the right panel of figure~\ref{fig:1gap-combined-results}.
While the evolution of both gain and photon yield as a function of the electric field has a strong dependence on the gap spacing, this dependence is strongly reduced when the photon yield and charge gain are measured against one another.
This behavior is consistent with the picture introduced in section~\ref{subsec:ArCF4}: in CF$_4$-based gas mixtures the scintillation yield is intrinsically coupled to electron avalanche.
Changing the gap spacing primarily changes the electron avalanche achieved at a given field, while the photon yield is more directly correlated with the achieved electron gain.
Above the low-gain region, the absolute photon yield follows approximately a power-law of the charge gain.
The dependence appears to differ slightly between the different spacings investigated: closer to linear in the case of the larger spacings and slightly sublinear in the case of the 1.6~mm gap.
This deviation from linearity as a function of gain also appears in the relative photon yield.
A likely contribution is photon feedback: feedback-emitted electrons contribute to the charge signal while part of their associated light appears in the delayed signal tail that is excluded from the photon yield, as detailed in section~\ref{subsec:SigProc}.

\subsection{Two-gap operation}
\label{subsec:2Gap}

As described in section~\ref{subsec:Det}, the amplification setup consists of two consecutive gaps, A1-A2 and A2-A3, with the final amplified charge collected on A3.

\begin{figure}[htbp]
    \centering
    \begin{subfigure}{0.48\linewidth}
        \includegraphics[width=\linewidth]{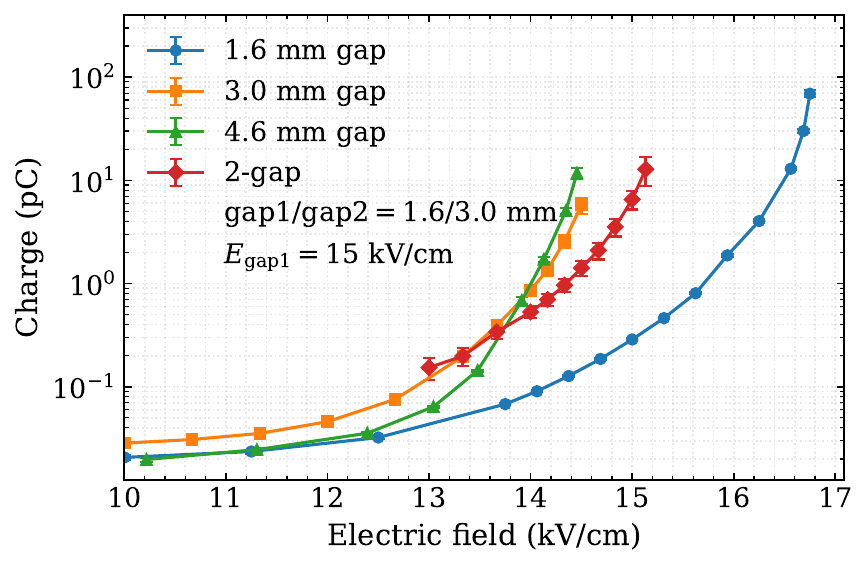}
        \caption{}
    \end{subfigure}
    \hfill
    \begin{subfigure}{0.48\linewidth}
        \includegraphics[width=\linewidth]{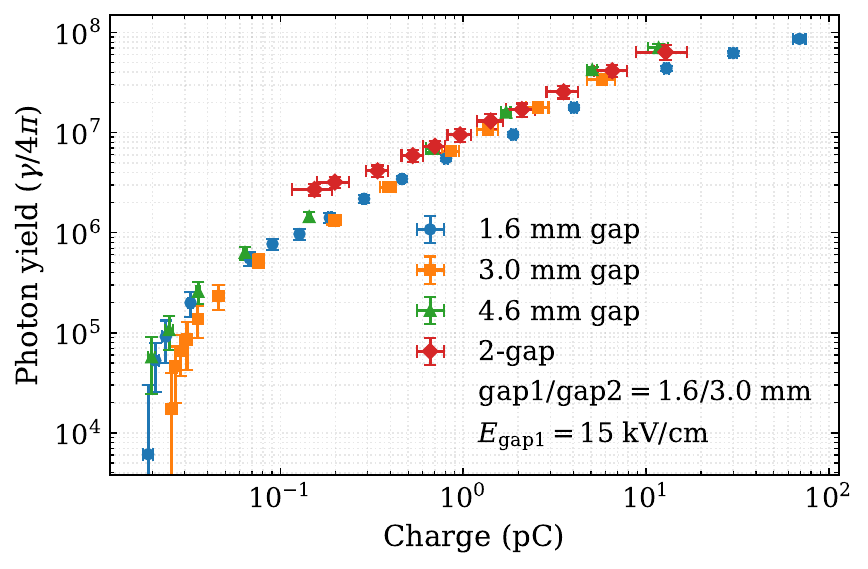}
        \caption{}
    \end{subfigure}
    \caption{Comparison of the single-gap and two-gap measurements. (a) Charge collected as a function of the electric field across the last gap. For the two-gap configuration the first gap was held at 15~kV/cm. (b) Inferred photon yield ($\gamma/4\pi$) as a function of the collected charge.}
    \label{fig:2gap-results}
\end{figure}

The first and second gap spacings were fixed to 1.6~mm and 3.0~mm, respectively.
Initial attempts to operate with a very high amplification through the first gap were unsuccessful: electrical instabilities and high voltage breakdowns occurred when transferring the charge into the second gap.
A moderate amplification on the first gap allowed for more stability across the second gap operating parameter space.
Under these conditions, the first gap operated under a field of 15~kV/cm, corresponding to 2.4~kV across 1.6~mm.
According to the single-gap measurements shown in figure~\ref{fig:1gap-results}, this corresponds to an effective charge gain of about 10 in the first stage.
The second-gap field was then scanned while the first-gap field was kept fixed.
The data were taken, processed, and analyzed following the same procedure as for the single-gap measurements, with the photon yield calculation adapted as described in section~\ref{subsec:Meas}.

Figure~\ref{fig:2gap-results} compares the two-gap measurements with the single-gap results.
For this comparison, instead of using the charge gain we used the actual charge collected on the final anode (A2 in single-gap and A3 in two-gap) to represent the result.
This is done in order to avoid ambiguities in defining the charge gain when comparing single-gap and two-gap results.

The results follow the same trends observed in single-gap operation.
In particular, the photon yield as a function of the amplified charge, displayed on the right panel of figure~\ref{fig:2gap-results}, overlaps neatly with the single-gap data.
This confirms further the previous observation that the amplification region geometry loosely affects the evolution of photon yield as a function of the amplified charge, which leads to the conclusion that, for the geometries evaluated, the factor limiting the photon throughput is essentially the achieved charge gain.

\subsection{Preliminary optical imaging}
\label{subsec:Track}

\begin{figure}[htbp]
    \centering
    \begin{subfigure}{0.48\linewidth}
        \includegraphics[width=\linewidth]{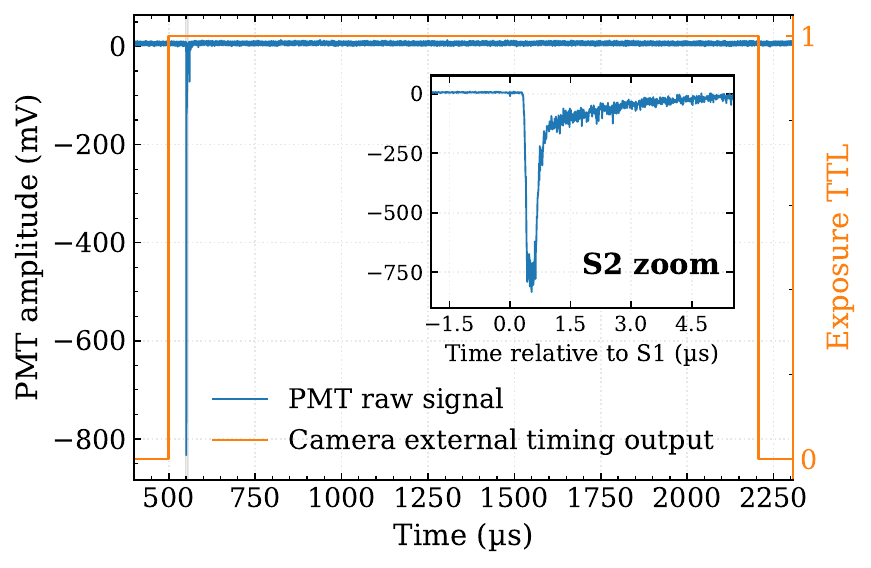}
        \caption{}
    \end{subfigure}
    \hfill
    \begin{subfigure}{0.48\linewidth}
        \includegraphics[width=\linewidth]{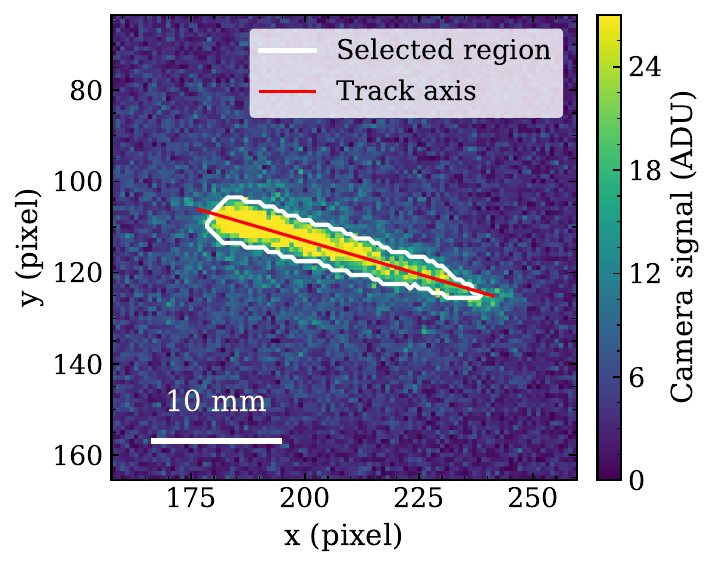}
        \caption{}
    \end{subfigure}
    \caption{Demonstration of synchronized track recording by the camera-PMT system and subsequent track identification. (a) PMT signal recording in coincidence with the camera global exposure timing output. A zoom on the avalanche signal recorded by the PMT is embedded where the time axis reference is the identified S1 signal. (b) Corresponding camera frame, cropped and rebinned 8 by 8, and containing the avalanche light measured by the PMT. The track selection contour is shown in white, while the track axis, obtained by principal component analysis (PCA), is shown in red.}
    \label{fig:cam-results}
\end{figure}

For the preliminary imaging studies, a Hamamatsu ORCA-Quest~2 qCMOS camera (C15550-22UP, Hamamatsu Photonics K.K., Japan) equipped with an ultra-fast lens (25~mm focal length, f/0.85 speed, model EHD-25085/C) was installed behind the fused-silica viewport and focused on the amplification plane, following the schematic shown in figure~\ref{fig:tpc_schematic}.
In this configuration the recorded frames correspond to a 2D projection of the avalanche light produced in the two-gap stage.
The camera and the PMT were synchronized by sending the camera global exposure timing output (TTL signal) to the oscilloscope, together with the PMT anode signal.
The camera TTL signal was used to trigger the oscilloscope, ensuring a one-to-one correspondence between the recorded camera frames and the recorded waveforms.
Because the present setup relied on a rolling-shutter camera and USB-based readout, the camera was operated in free-running mode and events were selected offline by requiring a single PMT pulse within the camera global exposure timing window to avoid pile-up and tracks cut by the rolling shutter.
An inherent limitation of the present setup is that the USB data link restricted the framerate and induced camera idle time.
This idle time limited the efficiency of the selection process.

In this configuration, matching camera frames and PMT waveforms were successfully recorded, demonstrating the feasibility of synchronized optical acquisition with the current prototype.
An example of a frame and the corresponding PMT waveform recorded on the oscilloscope is shown in figure~\ref{fig:cam-results}.
In this example, the detector was operated in the two-gap configuration producing the highest charge and light yield reported in section~\ref{subsec:2Gap}.
It should be noted that the delayed light signal tail, that we excluded from the S2 photon yield estimation, also contributes to the track brightness.
Although this delayed signal is not usable for direct time-resolved track imaging, it is nonetheless acquired by the camera during its exposure.

At this stage, this operation was used as a technical validation of the optical readout chain; a quantitative reconstruction study is left for future work since the current readout scheme still suffers from event loss and limited acquisition efficiency.

\section{Discussion and outlook}
\label{sec:Final}

An OTPC prototype was built and operated successfully with alpha particles, achieving an effective charge gain of the order of 1000 (corresponding to a gain of 560 when normalized to the primary ionization charge corrected for the Shockley-Ramo theorem); it is significantly higher than that reported in a previous and similar work~\cite{Fraga2001}, where a gain of about 30 was reached.
The results regarding amplified signals suggest that, for Ar:CF$_4$ at 1~bar, the maximum photon throughput is primarily set by the maximum charge gain, as the photon yield follows roughly the same dependence on charge for the parallel meshes configurations investigated.
The maximum charge achieved is in the range of 10 to 60~pC (\num{6e7}~to \num{4e8}~$e^-$ across the entire alpha track).
Electron transport measurements were also performed with the prototype and will be reported separately.

Future work, in view of future operation at SARAF-II, will explore the behavior of the detector for other gas mixtures.
At atmospheric pressure we were limited by the attainable operating voltages necessary to reach sufficient amplification in gas mixtures richer in CF$_4$.
Reducing the pressure was not an option in this study because the PMT base was not able to sustain the required voltage at lower pressures.
Another reason is that reducing the pressure would eventually extend the range of forward-emitted alphas beyond the limits of the current drift volume.
Helium mixed with CF$_4$ was also off-limits due to the permeation of helium through PMT glass windows.
The aforementioned upcoming detector and its gas system have been designed to allow these studies to be carried out.
We are also planning on replacing the PMT with silicon photomultipliers (SiPMs), which are not pressure-sensitive while also being equally operable in a helium-rich atmosphere.

The final imaging solution has not yet been decided.
Upgrading to the CoaXPress data link will allow operation of the ORCA-Quest~2 in lossless mode in a similar fashion to what was demonstrated by the CYGNO and Migdal collaborations~\cite{Amaro2023,Araujo2023}.
We are however considering alternative options such as a Timepix-based camera which can provide event-driven 3D imaging of ionization tracks, similar to the ARIADNE experiment~\cite{Lowe2020}.
We also have identified novel asynchronous Single-Photon Avalanche Diode (SPAD) cameras as a promising albeit unproven optical readout solution for OTPCs.

Based on the results shown, we also suggest exploring a new geometric concept to enhance the total scintillation yield of an OTPC using a CF$_4$-based mixture beyond the charge limitation aforementioned: the electron-loss assisted secondary scintillation.
This would in fact resemble the implementation of the multi-step avalanche chamber~\cite{Breskin1979}: a first light-charge amplification stage would be followed by a small gap where electrons would be intentionally inefficiently transferred to the next stage, in order to enable another stage of maximum light-charge production.
This electron loss could also be assisted by tuning the attachment rate accordingly in-between the amplification stages.
The voltage requirements for operation at 1~bar appear prohibitive but they could be workable at lower pressure.

\acknowledgments
We thank Dr.~Volker Dangendorf for providing the vacuum vessel and gas system used in these studies.
We also thank Dr.~Itamar Israelashvili for helping to calibrate the Mass Flow Controllers.
We are grateful to Marina Abramov and Meir Shoa for their assistance in the construction of the field cage.
This research was supported by the Pazy Foundation, the Minerva Foundation with funding from the Federal German Ministry for Education and Research, and the Nella and Leon Benoziyo Center for High Energy Physics.
We also extend our gratitude to the Krenter-Perinot Center for High-Energy Particle Physics, the Shimon and Golde Picker–Weizmann Annual Grant, and the Sir Charles Clore Prize for their support.
Special thanks are extended to Martin Kushner Schnur for his invaluable contribution to this research.

\bibliographystyle{JHEP}
\bibliography{biblio}

@article{Felkai2026,
    author    = {R. Felkai and M. Borysova and L. Moleri and J. Pienaar and D. Vartsky and A. Breskin and I. Mor and L. Weissman and S. Bressler},
    title     = {Development of a Multi-Purpose Optical {TPC} for Neutron-Induced Reaction Studies at {SARAF}},
    journal   = {PoS},
    volume    = {TAUP2025},
    pages     = {423},
    year      = {2026},
    doi       = {10.22323/1.511.0423}
}

@article{Dupont2020,
    author    = {E. Dupont and M. Bossant and R. Capote and A. D. Carlson and Y. Danon and M. Fleming and Z. Ge and H. Harada and O. Iwamoto and N. Iwamoto and A. Kimura and A. J. Koning and C. Massimi and A. Negret and G. Noguere and A. Plompen and V. Pronyaev and G. Rimpault and S. Simakov and A. Stankovskiy and W. Sun and A. Trkov and H. Wu and K. Yokoyama and the {NEA WPEC Expert Group on the High Priority Request List for Nuclear Data}},
    title     = {{HPRL}---International cooperation to identify and monitor priority nuclear data needs for nuclear applications},
    journal   = {EPJ Web Conf.},
    volume    = {239},
    pages     = {15005},
    year      = {2020},
    doi       = {10.1051/epjconf/202023915005}
}

@article{Mardor2022,
    author    = {Israel Mardor},
    title     = {A New Probe to the High-Intensity Frontier: Soreq Applied Research Accelerator Facility ({SARAF})},
    journal   = {Nucl. Phys. News},
    volume    = {32},
    number    = {4},
    pages     = {5--11},
    year      = {2022},
    doi       = {10.1080/10619127.2022.2100154}
}

@article{Hilke2010,
    author    = {H. J. Hilke},
    title     = {Time projection chambers},
    journal   = {Rep. Prog. Phys.},
    volume    = {73},
    number    = {11},
    pages     = {116201},
    year      = {2010},
    doi       = {10.1088/0034-4885/73/11/116201}
}

@book{Peisert1984,
    author    = {Anna Peisert and Fabio Sauli},
    title     = {Drift and Diffusion of Electrons in Gases},
    publisher = {CERN},
    address   = {Geneva},
    series    = {CERN Yellow Reports: Monographs},
    year      = {1984},
    doi       = {10.5170/CERN-1984-008}
}

@article{Christophorou1979,
    author    = {L. G. Christophorou and D. L. McCorkle and D. V. Maxey and J. G. Carter},
    title     = {Fast gas mixtures for gas-filled particle detectors},
    journal   = {Nucl. Instrum. Methods},
    volume    = {163},
    number    = {1},
    pages     = {141--149},
    year      = {1979},
    doi       = {10.1016/0029-554X(79)90042-9}
}

@article{Datskos1992,
    author    = {P. G. Datskos and J. G. Carter and L. G. Christophorou},
    title     = {Ionization coefficients in selected gas mixtures of interest to particle detectors},
    journal   = {J. Appl. Phys.},
    volume    = {71},
    number    = {1},
    pages     = {15--21},
    year      = {1992},
    doi       = {10.1063/1.350730}
}

@article{Anderson1992,
    author    = {W. S. Anderson and J. C. Armitage and E. Dunn and J. G. Heinrich and C. Lu and K. T. McDonald and J. Weckel and Y. Zhu},
    title     = {Electron attachment, effective ionization coefficient, and electron drift velocity for {CF$_4$} gas mixtures},
    journal   = {Nucl. Instrum. Methods Phys. Res. A},
    volume    = {323},
    number    = {1},
    pages     = {273--279},
    year      = {1992},
    doi       = {10.1016/0168-9002(92)90302-K}
}

@article{Pansky1995,
    author    = {A. Pansky and A. Breskin and A. Buzulutskov and R. Chechik and V. Elkind and J. Va'vra},
    title     = {{The scintillation of CF$_4$ and its relevance to detection science}},
    journal   = {Nucl. Instrum. Methods Phys. Res. A},
    volume    = {354},
    number    = {2},
    pages     = {262--269},
    year      = {1995},
    doi       = {10.1016/0168-9002(94)01064-1}
}

@article{Monteiro2007,
    author    = {C. M. B. Monteiro and L. M. P. Fernandes and J. A. M. Lopes and L. C. C. Coelho and J. F. C. A. Veloso and J. M. F. dos Santos and K. Giboni and E. Aprile},
    title     = {Secondary scintillation yield in pure xenon},
    journal   = {J. Instrum.},
    volume    = {2},
    number    = {05},
    pages     = {P05001},
    year      = {2007},
    doi       = {10.1088/1748-0221/2/05/P05001}
}

@article{Monteiro2008,
    author    = {C. M. B. Monteiro and J. A. M. Lopes and J. F. C. A. Veloso and J. M. F. {dos Santos}},
    title     = {Secondary scintillation yield in pure argon},
    journal   = {Phys. Lett. B},
    volume    = {668},
    number    = {3},
    pages     = {167--170},
    year      = {2008},
    doi       = {10.1016/j.physletb.2008.08.030}
}

@article{Mano2022,
    author    = {R. D. P. Mano and C. A. O. Henriques and F. D. Amaro and C. M. B. Monteiro},
    title     = {Secondary scintillation yield in pure krypton},
    journal   = {Phys. Lett. B},
    volume    = {824},
    pages     = {136782},
    year      = {2022},
    doi       = {10.1016/j.physletb.2021.136782}
}

@article{Araujo2023,
    author       = {H. M. Araújo and S. N. Balashov and J. E. Borg and F. M. Brunbauer and C. Cazzaniga and C. D. Frost and F. Garcia and A. C. Kaboth and M. Kastriotou and I. Katsioulas and A. Khazov and H. Kraus and V. A. Kudryavtsev and S. Lilley and A. Lindote and D. Loomba and M. I. Lopes and E. Lopez Asamar and P. Luna Dapica and P. A. Majewski and T. Marley and C. McCabe and A. F. Mills and M. Nakhostin and T. Neep and F. Neves and K. Nikolopoulos and E. Oliveri and L. Ropelewski and E. Tilly and V. N. Solovov and T. J. Sumner and J. Tarrant and R. Turnley and M. G. D. {van der Grinten} and R. Veenhof},
    title        = {The {MIGDAL} experiment: Measuring a rare atomic process to aid the search for dark matter},
    journal      = {Astropart. Phys.},
    volume       = {151},
    year         = {2023},
    pages        = {102853},
    doi          = {10.1016/j.astropartphys.2023.102853}
}

@article{Amaro2023,
    author       = {Fernando Domingues Amaro and Elisabetta Baracchini and Luigi Benussi and Stefano Bianco and Cesidio Capoccia and Michele Caponero and Danilo Santos Cardoso and Gianluca Cavoto and André Cortez and Igor Abritta Costa and Emiliano Dané and Giorgio Dho and Flaminia {Di Giambattista} and Emanuele {Di Marco} and Giulia D'Imperio and Francesco Iacoangeli and Herman Pessoa Lima and Guilherme Sebastiao Pinheiro Lopes and Giovanni Maccarrone and Rui Daniel Passos Mano and Robert Renz Marcelo Gregorio and David José Gaspar Marques and Giovanni Mazzitelli and Alasdair Gregor McLean and Andrea Messina and Cristina Maria Bernardes Monteiro and Rafael Antunes Nobrega and Igor Fonseca Pains and Emiliano Paoletti and Luciano Passamonti and Sandro Pelosi and Fabrizio Petrucci and Stefano Piacentini and Davide Piccolo and Daniele Pierluigi and Davide Pinci and Atul Prajapati and Francesco Renga and Rita Joanna {da Cruz Roque} and Filippo Rosatelli and Alessandro Russo and Joaquim Marques Ferreira {dos Santos} and Giovanna Saviano and Neil John Curwen Spooner and Roberto Tesauro and Sandro Tommasini and Samuele Torelli},
    title        = {The {CYGNO} experiment, a directional detector for direct dark matter searches},
    journal      = {Nucl. Instrum. Methods Phys. Res. A},
    volume       = {1054},
    year         = {2023},
    pages        = {168325},
    doi          = {10.1016/j.nima.2023.168325}
}

@article{LXCat2017,
    author    = {L. C. Pitchford and L. L. Alves and K. Bartschat and S. F. Biagi and M.-C. Bordage and I. Bray and C. E. Brion and M. J. Brunger and L. Campbell and A. Chachereau and B. Chaudhury and L. G. Christophorou and E. Carbone and N. A. Dyatko and C. M. Franck and D. V. Fursa and R. K. Gangwar and V. Guerra and P. Haefliger and G. J. M. Hagelaar and A. Hoesl and Y. Itikawa and I. V. Kochetov and R. P. McEachran and W. L. Morgan and A. P. Napartovich and V. Puech and M. Rabie and L. Sharma and R. Srivastava and A. D. Stauffer and J. Tennyson and J. de Urquijo and J. van Dijk and L. A. Viehland and M. C. Zammit and O. Zatsarinny and S. Pancheshnyi},
    title     = {{LXCat}: an Open-Access, Web-Based Platform for Data Needed for Modeling Low Temperature Plasmas},
    journal   = {Plasma Process. Polym.},
    volume    = {14},
    number    = {1--2},
    pages     = {1600098},
    year      = {2017},
    doi       = {10.1002/ppap.201600098}
}

@article{Bordage1996,
    author    = {M.-C. Bordage and P. S{\'e}gur and A. Chouki},
    title     = {Determination of a set of electron impact cross sections in tetrafluoromethane consistent with experimental determination of swarm parameters},
    journal   = {J. Appl. Phys.},
    volume    = {80},
    number    = {3},
    pages     = {1325--1336},
    year      = {1996},
    doi       = {10.1063/1.362931}
}

@article{Bordage1999,
    author    = {M.-C. Bordage and P. S{\'e}gur and L. G. Christophorou and J. K. Olthoff},
    title     = {Boltzmann analysis of electron swarm parameters in {CF$_4$} using independently assessed electron-collision cross sections},
    journal   = {J. Appl. Phys.},
    volume    = {86},
    number    = {7},
    pages     = {3558--3566},
    year      = {1999},
    doi       = {10.1063/1.371258}
}

@article{Fraga2001,
    author    = {M. M. R. Fraga and C. C. Bueno and J. A. C. Goncalves and F. A. F. Fraga and R. Ferreira Marques and A. J. P. L. Policarpo},
    title     = {Pressure dependence of secondary {NIR} scintillation in {Ar} and {Ar}/{CF$_4$}},
    journal   = {IEEE Trans. Nucl. Sci.},
    volume    = {48},
    number    = {3},
    pages     = {330--335},
    year      = {2001},
    doi       = {10.1109/23.940075}
}

@article{Fraga2003,
    author  = {M. M. F. R. Fraga and F. A. F. Fraga and S. T. G. Fetal and L. M. S. Margato and R. Ferreira Marques and A. J. P. L. Policarpo},
    title   = {{The {GEM} scintillation in {He}--{CF$_4$}, {Ar}--{CF$_4$}, {Ar}--{TEA} and {Xe}--{TEA} mixtures}},
    journal = {Nucl. Instrum. Methods Phys. Res. A},
    volume  = {504},
    number  = {1},
    pages   = {88--92},
    year    = {2003},
    doi     = {10.1016/S0168-9002(03)00758-7}
}

@article{Lowe2020,
    author    = {Adam Lowe and Krishanu Majumdar and Konstantinos Mavrokoridis and Barney Philippou and Adam Roberts and Christos Touramanis and Jared Vann},
    title     = {Optical Readout of the {ARIADNE} {LArTPC} Using a {Timepix3}-Based Camera},
    journal   = {Instruments},
    volume    = {4},
    number    = {4},
    pages      = {35},
    year      = {2020},
    doi       = {10.3390/instruments4040035}
}

@article{Gai2010,
    author    = {M. Gai and M. W. Ahmed and S. C. Stave and W. R. Zimmerman and A. Breskin and B. Bromberger and R. Chechik and V. Dangendorf and Th. Delbar and R. H. France III and S. S. Henshaw and T. J. Kading and P. P. Martel and J. E. R. McDonald and P.-N. Seo and K. Tittelmeier and H. R. Weller and A. H. Young},
    title     = {An optical readout {TPC} ({O-TPC}) for studies in nuclear astrophysics with gamma-ray beams at {HI$\gamma$S}},
    journal   = {J. Instrum.},
    volume    = {5},
    number    = {12},
    pages     = {P12004},
    year      = {2010},
    doi       = {10.1088/1748-0221/5/12/P12004}
}

@article{Gai2020,
    author    = {M. Gai and D. Schweitzer and S. R. Stern and A. H. Young and R. Smith and M. Cwiok and J. S. Bihalowicz and H. Czyrkowski and R. Dabrowski and W. Dominik and A. Fijalkowska and Z. Janas and L. Janiak and A. Korgul and T. Matulewicz and C. Mazzocchi and M. Pf{\"u}tzner and M. Zaremba and D. Balabanski and I. Gheorghe and C. Matei and O. Tesileanu and N. V. Zamfir and M. W. Ahmed and S. S. Henshaw and C. R. Howell and J. M. Mueller and L. S. Myers and S. Stave and C. Sun and H. R. Weller and Y. K. Wu and A. Breskin and V. Dangendorf and K. Tittelmeier and M. Freer},
    title     = {{Time Projection Chamber} ({TPC}) detectors for nuclear astrophysics studies with gamma beams},
    journal   = {Nucl. Instrum. Methods Phys. Res. A},
    volume    = {954},
    pages     = {161779},
    year      = {2020},
    doi       = {10.1016/j.nima.2019.01.006}
}

@article{Breskin1979,
    author    = {A. Breskin and G. Charpak and S. Majewski and G. Melchart and G. Petersen and F. Sauli},
    title     = {The multistep avalanche chamber: A new family of fast, high-rate particle detectors},
    journal   = {Nucl. Instrum. Methods},
    volume    = {161},
    number    = {1},
    pages     = {19--34},
    year      = {1979},
    doi       = {10.1016/0029-554X(79)90358-6}
}

@article{Breskin1990,
    author    = {A. Breskin and R. Chechik and D. Sauvage},
    title     = {A 3-stage gated {UV}-photon gaseous detector with optical imaging},
    journal   = {Nucl. Instrum. Methods Phys. Res. A},
    volume    = {286},
    number    = {1},
    pages     = {251--261},
    year      = {1990},
    doi       = {10.1016/0168-9002(90)90229-Y}
}

@article{Sauvage1989,
    author    = {D. Sauvage and A. Breskin and R. Chechik},
    title     = {A systematic study of the emission of light from electron avalanches in low-pressure {TEA} and {TMAE} gas mixtures},
    journal   = {Nucl. Instrum. Methods Phys. Res. A},
    volume    = {275},
    number    = {2},
    pages     = {351--363},
    year      = {1989},
    doi       = {10.1016/0168-9002(89)90709-2}
}

@article{Deisting2021,
    author    = {Alexander Deisting and Abigail Victoria Waldron and Edward Atkin and Gary Barker and Anastasia Basharina-Freshville and Christopher Betancourt and Steven Boyd and Dominic Brailsford and Zachary Chen-Wishart and Linda Cremonesi and Adriana Dias and Patrick Dunne and Jennifer Haigh and Philip Hamacher-Baumann and Sebastian Jones and Asher Kaboth and Alexander Korzenev and William Ma and Philippe Mermod and Maria Mironova and Jocelyn Monroe and Ryan Nichol and Toby Nonnenmacher and Jaroslaw Nowak and William Parker and Harrison Ritchie-Yates and Stefan Roth and Ruben Saakyan and Nicola Serra and Yuri Shitov and Jochen Steinmann and Adam Tarrant and Melissa Uchida and Sammy Valder and Mark Ward and Morgan O. Wascko},
    title     = {Commissioning of a High Pressure Time Projection Chamber with Optical Readout},
    journal   = {Instruments},
    volume    = {5},
    number    = {2},
    pages     = {22},
    year      = {2021},
    doi       = {10.3390/instruments5020022}
}

@article{Morozov2011,
    author    = {A. Morozov and M. M. F. R. Fraga and L. Pereira and L. M. S. Margato and S. T. G. Fetal and B. Guerard and G. Manzin and F. A. F. Fraga},
    title     = {Effect of the electric field on the primary scintillation from {CF$_4$}},
    journal   = {Nucl. Instrum. Methods Phys. Res. A},
    volume    = {628},
    number    = {1},
    pages     = {360--363},
    year      = {2011},
    doi       = {10.1016/j.nima.2010.07.001}
}

@phdthesis{Brunbauer2018,
    author = {Florian M. Brunbauer},
    title  = {Applications of Gas Scintillation Properties in Optically Read Out {GEM}-Based Detectors},
    school = {Technische Universit{\"a}t Wien},
    year   = {2018},
    month  = jul
}

@article{Amedo2023,
    author    = {P. Amedo and D. Gonz{\'a}lez-D{\'i}az and F. M. Brunbauer and D. J. Fern{\'a}ndez-Posada and E. Oliveri and L. Ropelewski},
    title     = {Observation of strong wavelength-shifting in the argon-tetrafluoromethane system},
    journal   = {Front. Detect. Sci. Technol.},
    volume    = {1},
    pages     = {1282854},
    year      = {2023},
    doi       = {10.3389/fdest.2023.1282854}
}

@article{Brunbauer2025,
    author    = {F. M. Brunbauer and P. Amedo and K. J. Fl{\"o}thner and D. Gonzalez Diaz and D. Janssens and S. Leardini and M. Lisowska and H. M{\"u}ller and E. Oliveri and G. Orlandini and D. Pfeiffer and L. Ropelewski and F. Sauli and J. Samarati and L. Scharenberg and M. van Stenis and R. Veenhof},
    title     = {Primary and secondary scintillation of {CF$_4$}-based mixtures in low-pressure gaseous detectors},
    journal   = {Front. Detect. Sci. Technol.},
    volume    = {3},
    pages     = {1561739},
    year      = {2025},
    doi       = {10.3389/fdest.2025.1561739}
}

@article{Miernik2007,
    author    = {K. Miernik and W. Dominik and H. Czyrkowski and R. D{\k a}browski and A. Fomitchev and M. Golovkov and Z. Janas and W. Ku{\'s}mierz and M. Pf{\"u}tzner and A. Rodin and S. Stepantsov and R. Slepniev and G. M. Ter-Akopian and R. Wolski},
    title     = {Optical Time Projection Chamber for imaging nuclear decays},
    journal   = {Nucl. Instrum. Methods Phys. Res. A},
    volume    = {581},
    number    = {1},
    pages     = {194--197},
    year      = {2007},
    doi       = {10.1016/j.nima.2007.07.076}
}

@article{He2001,
    author    = {Zhong He},
    title     = {Review of the {Shockley--Ramo} theorem and its application in semiconductor gamma-ray detectors},
    journal   = {Nucl. Instrum. Methods Phys. Res. A},
    volume    = {463},
    number    = {1},
    pages     = {250--267},
    year      = {2001},
    doi       = {10.1016/S0168-9002(01)00223-6}
}

@article{SRIM2010,
    author    = {J. F. Ziegler and M. D. Ziegler and J. P. Biersack},
    title     = {{SRIM}---The Stopping and Range of Ions in Matter (2010)},
    journal   = {Nucl. Instrum. Methods Phys. Res. B},
    volume    = {268},
    number    = {11},
    pages     = {1818--1823},
    year      = {2010},
    doi       = {10.1016/j.nimb.2010.02.091}
}
\end{document}